\documentclass[twocolumn,english,aps,prb,floatfix,showpacs]{revtex4}
\usepackage[T1]{fontenc}
\usepackage[latin1]{inputenc}
\usepackage{babel}
\usepackage{graphics}

\makeatletter
\usepackage{graphicx}
\usepackage{amsmath}
\usepackage[T1]{fontenc}
\usepackage[latin1]{inputenc}
\usepackage{times}
\usepackage[normalem]{ulem}
\usepackage{threeparttable}

  \makeatletter
  \def\@dotsep{4.5}
  \makeatother

\newlength{\myVSpace}
\makeatother
\begin{document}

\title{An Improved Real--Space Genetic Algorithm for Crystal Structure and
Polymorph Prediction}

\author{N.L. Abraham%
\thanks{Present address: Centre for Atmospheric Science, University of Cambridge,
Lensfield Road, Cambridge, CB2 1EW, UK
}}

\author{M.I.J. Probert%
\thanks{Author to whom correspondence should be addressed
}}

\affiliation{Department of Physics, University of York, Heslington, York, YO10
5DD, United Kingdom}


\begin{abstract}
Existing Genetic Algorithms for crystal structure and polymorph prediction
can suffer from stagnation during evolution, with a consequent loss
of efficiency and accuracy. An improved Genetic Algorithm (GA) is
introduced herein which penalizes similar structures and so enhances
structural diversity in the population at each generation. This is
shown to improve the quality of results found for the theoretical
prediction of simple model crystal structures. In particular, this
method is demonstrated to find three new zero--temperature phases
of the Dzugutov potential that have not been previously reported. 
\end{abstract}
\maketitle

\section{Introduction}

Genetic algorithms (GAs) are emerging as a useful tool in the theoretical
prediction of crystal structures (see \citet[and references therein]{abraham:224104})
\citep{trimarchi:104113,GlassOH06,briggs:195415}. During a GA calculation
it is possible that the system will \textit{stagnate}. When stagnation
occurs, one or more local minima dominate the search and the method
is unable to find the global minimum solution. In this communication
we improve the convergence to the global minimum solution of the CASTEP--GA
\citep{abraham:224104} through the use of a fitness function which
is able to differentiate structures during the course of a GA minimization.

Binary--encoded GAs such as the method of \citet{HartBWZ05, BlumHWZ05}
are able to directly compare the binary strings that make up their
population members and determine if two populations members are the
same. In this way it is possible to remove any highly prevalent local
minimum from the population, and prevent its creation in future mating
operations. While this method is not possible in the frame--work of
the CASTEP--GA, we have developed an alternative approach that significantly
reduces the stagnation rate and also forces the system to explore
new minima. This alternative approach is also broadly transferable
to a wide range of other GAs.

\section{Method}

Our GA method \citep{abraham:224104} is a real--space encoded technique
for crystal structure prediction which takes advantage of the periodicity
of the simulation supercell to improve the speed and accuracy of convergence
to the global minimum free-energy crystal structure. There is a \textit{population}
of structures (or \textit{members}) which are bred together to produce
new members, such that with each subsequent \textit{generation} the
population evolves in an attempt to determine the global minimum structure.
The \textit{fitness function} of the GA is used to determine how good
({}``fit'') a structure is and this is then used to weight the probability
of survival of that structure and its probability to produce offspring.

While this method has been very successful in the past, we wanted
to reduce the stagnation rate and thereby improve the quality of the
solutions produced during a GA structure search. Since this is a real--space
based approach it is not possible to directly compare the atomic co--ordinates
of two population members to determine if they are the same structure.
In our previous work \citep{abraham:224104}, the enthalpy of the structure
was used to calculate the fitness. In this work, we propose augmenting
this fitness function with an additional function which is able to
determine the similarity of two structures. We shall illustrate the
effectiveness of this new approach by first studying the Lennard--Jones
crystals for comparison with our previous results and then the high
pressure phases of the Dzugutov potential \citep{PhysRevLett.70.2924}.

The enthalpy--based fitness function is \begin{equation}
\label{eq:fit1}
f_{i}=\frac{\left[ 1-\tanh (2\rho _{i}-1)\right] }{2}
\end{equation}
 with the variable \( \rho _{i} \) being defined by \begin{equation}
\label{eq:fit2}
\rho _{i}=\frac{V_{i}-V_{min}}{V_{max}-V_{min}}
\end{equation}
 where \( V_{max} \) is the enthalpy of the highest enthalpy member
of the population, \( V_{min} \) is the enthalpy of the lowest enthalpy
member and \( V_{i} \) is the enthalpy of the member \( i \) being
considered. The fitness of each member \( i \) is \( f_{i} \) and
this is a function which varies between zero and one. Population members
with a fitness close to zero are less {}``fit'', and members with
a fitness close to one are more {}``fit''. Population members are
then selected (using roulette--wheel selection) for reproduction or
are removed from the population based on this fitness value.

This should mean that only fit members are selected to remain in the
population, or are allowed to breed (\textit{crossover}). It is often
very likely that during the course of a calculation multiple copies
of population members are made. In a bit--string represented GA duplicate
members are very easy to spot but in a real--space encoded GA it is
very hard to tell if two members are the same during the course of
a calculation, since the crystal structure may be orientated or translated
in any way within the simulation cell (due to use of periodic boundary
conditions). This is even harder if the simulation cell parameters
are also allowed to vary during the course of a calculation.

Hence we need a simple measure of structural similarity so that we
can detect when duplicate structures exist within a population. Whilst
this is encouraging from the point of view of ultimate structural
convergence, in the early stages of the GA minimization we want to
ensure as much structural diversity as possible to enable a broad
search of possible solutions and so we want to penalise similar structures.

Since we are using this routine to differentiate between like and
unlike structures, rather than any form of comprehensive structural
analysis, we can simplify this comparison somewhat. If we are performing
a calculation in which we allow the number of atoms to vary, then
we can make an educated guess that two structures with different numbers
of atoms are different (or rather in this case any offspring produced
in the crossover procedure will have a greater number of degrees of
freedom to explore the potential energy surface), so we will have
no need to compare these structures. We also do not need to compare
each structure with all other structures, since we are merely trying
to prevent stagnation rather than give a definitive structural comparison,
and so we can simply compare all structures with the minimum enthalpy
structure which has the same number of atoms as itself that exist
in the current generation. We will define a comparison function between
structures which returns zero if the structures are the same and one
if the structures are suitably dissimilar. We also want to keep the
fact that lower enthalpy structures are {}``better'' than higher
enthalpy ones, so we further weight the value that any given structure
has by the value of \( f_{i} \) of the fittest member in that {}``set''
which is made up of members with the same number of atoms. Here we
define our improved fitness function as \begin{equation}
\label{eqn. new_fit}
\, ^{j}\hspace {-2.5pt}f^{\prime }_{i}=\left( 1-\omega \right) \, ^{j}\hspace {-2.5pt}f_{i}+\omega \, ^{j}\hspace {-2.5pt}f_{fit}\star \left\{ \begin{array}{cc}
1 & i\equiv fit\\
R\left( \Lambda \left( k_{r}\right) \right)  & i\not \equiv fit
\end{array}\right. 
\end{equation}
where the left--superscript \( j \) above denotes comparing between
groups with the same number of atoms only, \( f_{i} \) is as defined
in equation \ref{eq:fit1}, \( w \) is a weighting value between
zero and one and \( R\left( \Lambda \left( k_{r}\right) \right)  \)
is a function which compares member \( i \) of the set of atoms \( j \)
with the fittest member in that set (as defined by equation \ref{eq:fit1}).
This means that the fitness of the fittest member of each group (\( \, ^{j}\hspace {-2.5pt}f_{fit} \))
will be unchanged from its enthalpy value, and all other values in
the group will be scaled accordingly. If the value of the fitness
weight, \( w \), is set to \( 1 \) then the maximum value of \( \, ^{j}\hspace {-2.5pt}f^{\prime }_{i} \)
that any member could have is the same value of the fittest member
of the group, \( \, ^{j}\hspace {-2.5pt}f_{fit} \). If \( w \) is
set to zero then this function reduces to that given in equation \ref{eq:fit1}.
The comparison function \( R\left( \Lambda \left( k_{r}\right) \right)  \)
is

\begin{equation}
\label{eq:comp1}
R\left( \Lambda \left( k_{r}\right) \right) =\frac{\sum _{k_{r}}|\Lambda ^{\prime }\left( k_{r}\right) -\Lambda \left( k_{r}\right) |}{\sum _{k_{r}}\Lambda ^{\prime }\left( k_{r}\right) }
\end{equation}
 where consideration of the spherically averaged scattering intensity
leads to \begin{eqnarray}
\Lambda \left( k_{r}\right)  & = & \Omega ^{2}\left[ N\sum _{n=1}^{N}\rho ^{\prime 2}\left( n\right) \right. +\label{eq:comp2} \\
 &  & \left. 2\sum _{n=1}^{N}\sum _{m>n}^{N}\rho ^{\prime 2}\left( n\right) \rho ^{\prime 2}\left( m\right) J_{0}\left( \sqrt{3}\pi k_{r}\left| \mathbf{r}_{n}-\mathbf{r}_{m}\right| \right) \right] \nonumber 
\end{eqnarray}
which is positive--definite (and is based on the Debye scattering
formula \citep{Debye15}) . In equation \ref{eq:comp2}, there are
\( N \) ions within the simulation cell which has a volume \( \Omega  \),
\( \rho ^{\prime }\left( n\right)  \) is the scattering factor of
ion \( n \) which has the atomic real--space co--ordinate of \( \mathbf{r}_{n} \),
and \( J_{0}\left( r\right)  \) is a Bessel function. The function
\( \Lambda \left( k_{r}\right)  \) of a population member \( i \)
of each group \( j \) is tested against the function \( \Lambda ^{\prime }\left( k_{r}\right)  \)
of the fittest member in the group \( j \) containing the same number
of atoms as member \( i \). Equation \ref{eq:comp1} is then used
to compare these two functions and returns a single number between
zero and one. In a variable--supercell calculation it is possible
for this function to become greater than one when the structures are
highly dissimilar, in which case we set the value of \( R\left( \Lambda \left( k_{r}\right) \right)  \)
to one.

\section{Results}

\begin{figure}
{\centering \resizebox*{0.9\columnwidth}{!}{\includegraphics{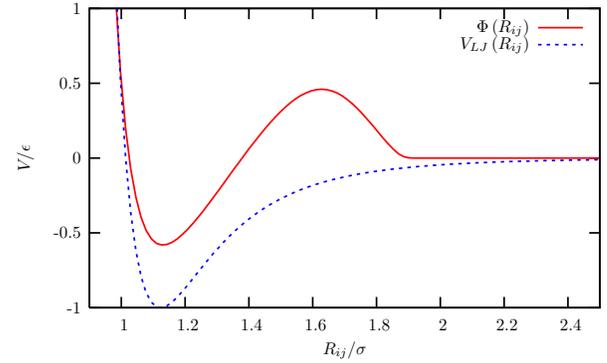}} \par}

\caption{(Color online) Comparision of the Lennard-Jones and Dzugutov pair-potentials.\label{fig:LJDZ}}
\end{figure}

The results presented here will use two different empirical potentials,
the Lennard--Jones potential \citep{StoddardF73,LennardJonesI25} and
the Dzugutov potential \citep{PhysRevLett.70.2924} which is defined
as\begin{equation}
\label{eq:dz}
\Phi \left( R_{ij}\right) =\Phi _{1}\left( R_{ij}\right) +\Phi _{2}\left( R_{ij}\right) 
\end{equation}
 where

\begin{eqnarray}
\Phi _{1} & = & \left\{ \begin{array}{lr}
A\left( R^{-m}_{ij}-B\right) \exp \left( \frac{c}{R_{ij}-a}\right)  & R_{ij}<a\\
0 & R_{ij}\geq a
\end{array}\right. \label{eq:dz0_1} \\
\Phi _{2} & = & \left\{ \begin{array}{lr}
B\exp \left( \frac{d}{R_{ij}-b}\right)  & \; \; \; \; \; \; \; \; \; \; \; \; \; \; \; \; \; \; \; R_{ij}<b\\
0 & R_{ij}\geq b
\end{array}\right. \label{eq:dz0_2} 
\end{eqnarray}
with the constants defined in table \ref{tab:dz}. A comparison of
these two potentials is shown in figure \ref{fig:LJDZ}. The Dzugutov
potential was originally formulated to simulate liquid systems, however
it has also been shown to have some interesting solid phases \citep{RothD00},
and can also be used to form quasi--crystals \citep{PhysRevLett.70.2924}.

\begin{table}
{\centering \begin{tabular}{ccccccc}
 \( m \)&
 \( A \)&
 \( c \)&
 \( a \)&
 \( B \)&
 \( d \)&
 \( b \)\\
\hline
16 &
 5.82 &
 1.1 &
 1.87 &
 1.28 &
 0.27 &
 1.94 \\
\end{tabular}\par}

\caption{Table of parameters used in the Dzugutov potential (equation \ref{eq:dz}).}

\label{tab:dz}
\end{table}

The Dzugutov potential is designed such that the force on, and energy
of, an atom moving within the potential go to zero at \( b/\sigma  \).
As is reported in \citet{RothD00} the Dzugutov potential has three
known stable phases at varying pressures: BCC, the \( \sigma  \)--phase
and FCC.

\subsection{Results from the Lennard-Jones potential}

\begin{figure}
{\centering \resizebox*{0.9\columnwidth}{!}{\includegraphics{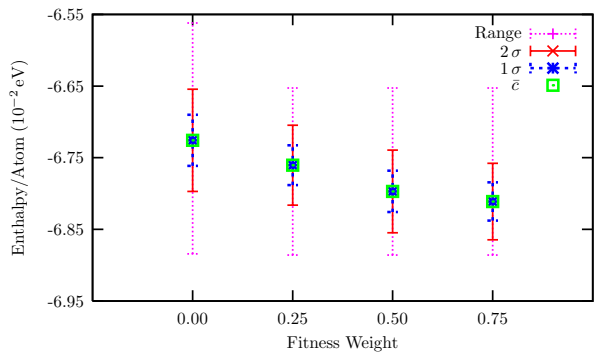}} \par}

\caption{(Color online) Summary of the enthalpies of the different Lennard-Jones
structures found for different fitness weights, which controls how
much the comparison factor is considered during selection for update
and crossover. The values for \protect\( w=0.0\protect \) are those
from \citet{abraham:224104}. All points are averaged over 15 independent
calculations.\label{fig:enth_sfac}}
\end{figure}

The use of the comparison factor in the selection procedure has a
marked effect on the quality of the results produced as seen in figure
\ref{fig:enth_sfac}. While the global minimum structure is hexagonal
close packed (HCP) this structure is nearly degenerate with the face-centred
cubic structure (FCC) (with an energy difference of less than \( 0.1\, \% \) \citep{KaneG40}).
There are also a number of other stacking--fault structures which
exist in-between FCC and HCP. The use of the comparison factor encourages
the system to explore and hence escape from these local minima and
find the HCP structure. With a fitness weight of \( w=0.75 \) finding
a HCP structure is much more likely.

\begin{figure}
{\centering \resizebox*{0.9\columnwidth}{!}{\includegraphics{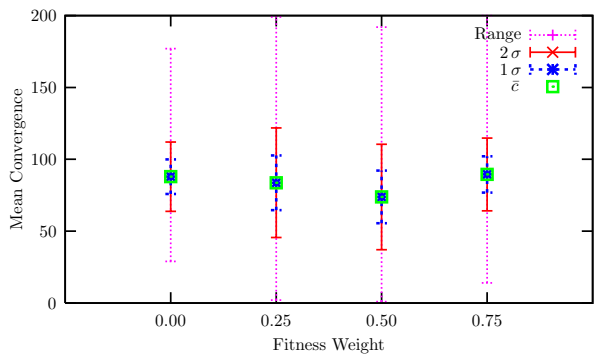}} \par}

\caption{(Color online) Summary of the convergence times for the results shown
in figure \ref{fig:enth_sfac}. The values for \protect\( w=0.0\protect \)
are those from \citet{abraham:224104}. All points are averaged over
15 independent calculations.\label{fig:conv_sfac}}
\end{figure}

The effect on convergence is interesting as seen in figure \ref{fig:conv_sfac}.
There is little increase in the mean number of generations required
for convergence, although there is a greater spread in the values.

\begin{figure}
{\centering \includegraphics*[scale=0.70]{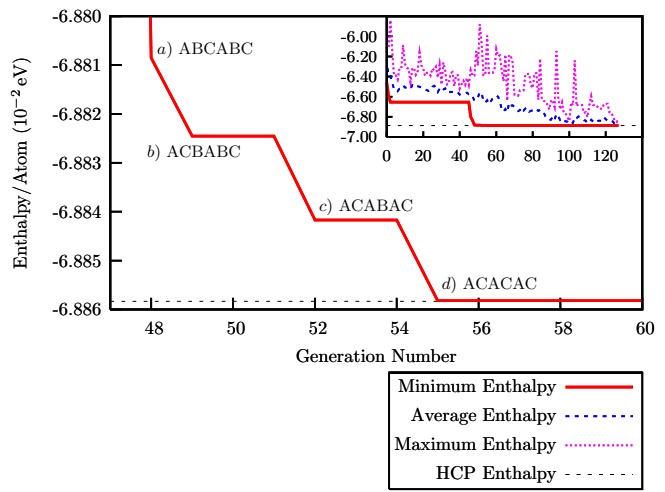}\par}

\caption{(Color online) Plot showing convergence to HCP minimum structure
for a Lennard-Jones calculation with \protect\( w=0.75\protect \).
The stacking patterns of the minimum enthalpy solutions are shown
next to their appearance during the course of the simulation. The
system converged to a HCP structure in 55 generations, and by the
\protect\( 127^{\mbox {\small th}}\protect \) generation all members
were the same. }

\label{fig:conv_sfac2}
\end{figure}

Figure \ref{fig:conv_sfac2} shows the results from a calculation
performed with \( w=0.75 \). We have included these results in particular
because it shows the system going from an FCC structure to a HCP structure
through two intermediate stacking--fault structures.

These results are summarized in table \ref{tab:sfac_strucs}. The
results from \( w=0.00 \) are those presented in \citet{abraham:224104}.

\begin{table}
{\centering \begin{tabular}{p{50pt}p{50pt}p{50pt}p{50pt}}
 \textbf{Fitness Weight}&
 \textbf{Pure HCP}&
 \textbf{Intermediate HCP--FCC}&
 \textbf{Pure FCC}\\
\hline
\( 0.00 \)&
 \( 0 \)&
 \( 6 \)&
 \( 0 \)\\
 \( 0.25 \)&
 \( 3 \)&
 \( 3 \)&
 \( 0 \)\\
 \( 0.50 \)&
 \( 3 \)&
 \( 6 \)&
 \( 0 \)\\
 \( 0.75 \)&
 \( 6 \)&
 \( 3 \)&
 \( 0 \)\\
\end{tabular}\par}

\caption{Table comparing the number of each ordered structure type of the
lowest enthalpy structure found (i.e. ignoring higher--enthalpy structures
found during the course of a GA minimization) for different values
of the fitness weighting factor \protect\( w\protect \). Numbers
given are out of a total of 15 calculations.}

{\centering \label{tab:sfac_strucs}\par}
\end{table}

\subsection{Results from the Dzugutov potential}

For results obtained using this potential an additional modification
was made to the GA in the crossover step. Previously the atom--number
could either be kept fixed or be allowed to vary in an unconstrained
manner. For these Dzugutov calculations a third option was added,
which is to allow the atom number to vary within an allowed percentage
of the original number of atoms within the simulation supercell. 

While this is not necessary in a fixed--cell size/shape calculation,
for a variable--cell calculation it is essential. Without this constraint
it would be possible for the number of atoms to keep decreasing with
the cell getting smaller and smaller until the minimum image convention
is violated, at which point the calculation will stop. It might also
allow a calculation to keep adding atoms at the crossover stage and
then allow the cell to grow to accommodate them. In this way the calculation
would increase in size and take a longer and longer time for each
minimization step. This percentage cut--off keeps the advantages of
a variable atom--number calculation without these problems.

\begin{figure}
{\centering \resizebox*{0.9\columnwidth}{!}{\includegraphics{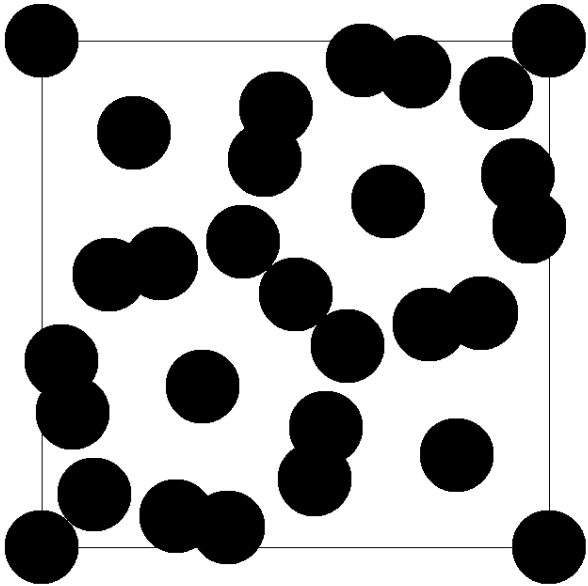}} \par}

\caption{The unit cell of the Dzugutov potential \protect\( \sigma \protect \)--phase
looking down the \protect\( \left[ 00\bar{1}\right] \protect \) direction.\label{fig:dz_s0}}
\end{figure}

It is already known that the Dzugutov \( \sigma  \)--phase has a
complicated 30--atom unit cell (see figure \ref{fig:dz_s0}) and so
all calculations had to have at least this many atoms. To prevent
any bias of the final results, we started each run with 62 atoms in
the unit cell and allowed the number of atoms to vary, in order to
have an unbiased search of a large enough phase space.

A summary of the Dzugatov results is given in table \ref{tab:dz_runsum}.
Calculations were performed at four pressures, 0\,MPa, 50\,MPa, 100\,MPa
and 150\,MPa which allows each of the three structures suggested by
\citet{RothD00} to be the most stable at at least one point during
the experiment.

\begin{table}
{\centering \begin{threeparttable}
\begin{tabular}{ r | l | r r r r r }
                            &                              & 
\multicolumn{5}{|c}{\textbf{Number of Each Phase Found}} \\
\hline
                           & {\textbf{\small Lowest}}              & 
{\textbf{\small Higher}}   &     &          &     & {\textbf{\small Lower}}  \\
\multicolumn{1}{c|}{\textbf{Pressure}} & {\small \textbf{Enthalpy}}            & 
{\textbf{\small Enthalpy}} &     &          &     & {\textbf{\small Enthalpy}} \\
\multicolumn{1}{c|}{\textbf{(MPa)}}    & {\small \textbf{Phase}} \tnote{\ddag} & 
{\textbf{\small Phase}}   & {\textbf{BCC}} & {\textbf{\(\sigma\)}} & {\textbf{FCC}} & 
{\textbf{\small Phase}} \tnote{\(\ast\)} \\
\hline
0   & BCC      & 13 & 8  & 1  & 0  & 0 \\
50  & BCC      & 2  & 16 & 1  & 0  & 3 \\
100 & \(\sigma\) & 1  & 9  & 11 & 0  & 1 \\
150 & FCC      & 1  & 0  & 0  & 15 & 6 \\
\end{tabular}
\begin{tablenotes}
\item[\ddag] Data taken from \citet{RothD00}.
\item[$\ast$] {Where the term ``Lower Enthalpy'' refers to having lower 
enthalpy than the phase in column 2.}  
\end{tablenotes}
\end{threeparttable}\par}

\caption{Summary of results for 62--atom variable--cell, constrained variable--atom--number
calculations. 22 independent GA calculations were performed at each
pressure.}

\label{tab:dz_runsum}
\end{table}

\begin{figure}
{\centering \resizebox*{0.9\columnwidth}{!}{\includegraphics{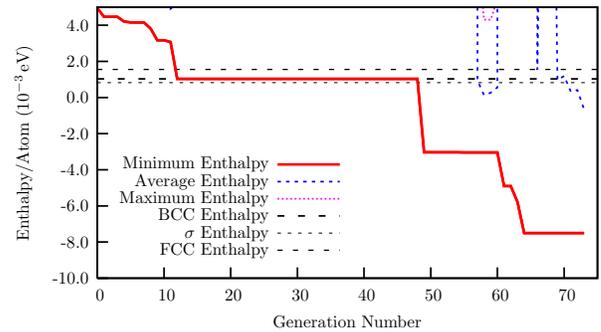}} \par}

\caption{(Color online) Convergence plot of a variable--atom, variable--cell
calculation, starting from 62 atoms. This gives rise to a previously
unknown phase (labeled as phase {}``\protect\( a\protect \)'' in
figure \ref{fig:dz_rad}). The inset shows the complete calculation.
The minimum--enthalpy structure found has 65--atoms and is shown in
figure \ref{fig:dz62_strB}.}

\label{fig:dz62_strA}
\end{figure}

\begin{figure}
{\centering \resizebox*{0.9\columnwidth}{!}{\includegraphics{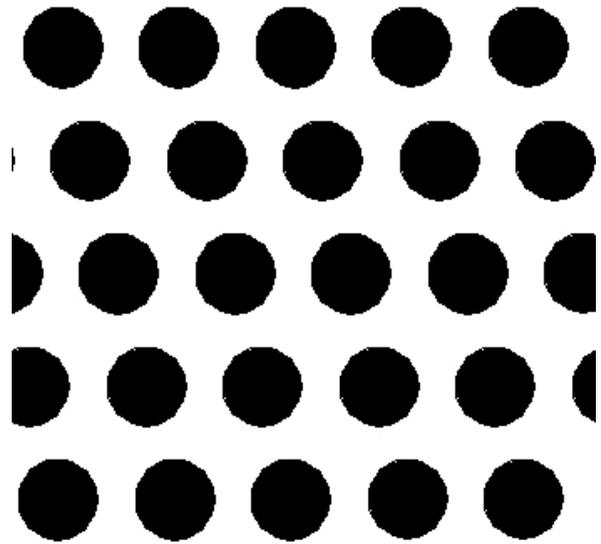}} \par}

\caption{The structure of the new phase {}``\protect\( a\protect \)'',
a 65--atom phase found in generation 64 of the calculation shown in
figure \ref{fig:dz62_strA}. }

\label{fig:dz62_strB}
\end{figure}

As can be seen in table \ref{tab:dz_runsum} a number of GA minimizations
found structures with a lower enthalpy than the previously reported
minimum enthalpy structure. In total three distinct new structures
were found. A plot showing the progress of a GA minimization down
to the new lowest--enthalpy structure found is shown in figure \ref{fig:dz62_strA}
with the structure itself shown in figure \ref{fig:dz62_strB}.

\begin{figure}
{\centering \resizebox*{0.9\columnwidth}{!}{\includegraphics{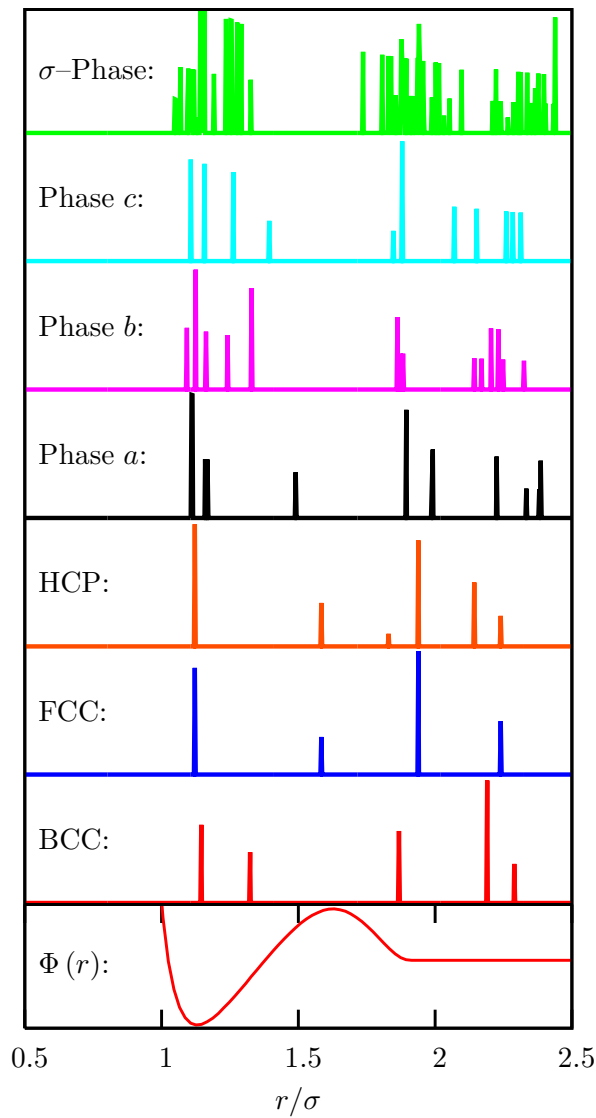}} \par}

\caption{(Color online) Comparison of the radial distribution function, \protect\( g\left( r\right) \protect \),
for the distinct lower--enthalpy structures found with BCC, FCC, HCP
and the \protect\( \sigma \protect \)--phase. The Dzugutov Potential
is also shown.}

\label{fig:dz_rad}
\end{figure}

A plot comparing the radial distribution functions of all the known
and unknown phases of the Dzugutov potential is shown in figure \ref{fig:dz_rad},
as well as the HCP phase. The three new phases found are all significantly
different from the established phases of this potential. Examination
of these phases suggests that the simulation cells correspond to primitive
cells and not supercells, but attempts to further characterize these
structures by space group have so far been unsuccessful. The atomic
co--ordinates of these structures are available online\cite{cmd_web}
as supplementary material.

\begin{figure}
{\centering \resizebox*{0.9\columnwidth}{!}{\includegraphics{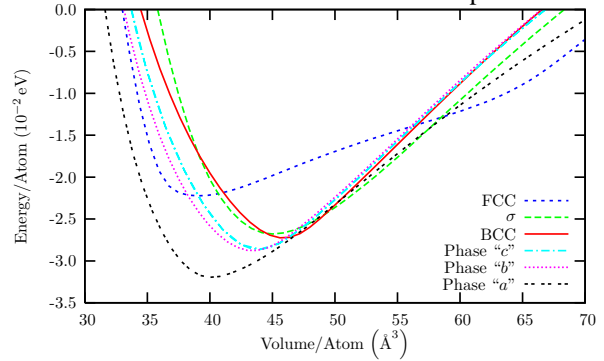}} \par}

\caption{(Color online) Energy--Volume curve for the Dzugutov potential showing
the three new phases calculated at zero pressure. The curves for the
\protect\( \sigma \protect \)--phase and structures {}``\protect\( a\protect \)'',
{}``\protect\( b\protect \)'' and {}``\protect\( c\protect \)''
were calculated assuming an isotropic expansion.}

\label{fig:dz_evcurve}
\end{figure}

A plot showing the energy--volume curves for the six phases of the
Dzugutov potential found in the course of this study is shown in figure
\ref{fig:dz_evcurve}. Phase {}``\( a \)'' is the most stable phase
at all positive pressures.

\section{Conclusions}

In this paper we have developed a novel fitness function that combines
a traditional approach to fitness based upon enthalpy, with a simple
structural comparison factor to find new, more stable crystal structures
within a GA for crystal structure prediction. This method penalises
the presence of similar structures within the population which prevents
the GA stagnating in some local minimum. The GA method itself was
also extended to allow both the simulation supercell and the number
of atoms within that supercell to vary. The number of atoms must only
be varied within fixed limits to prevent the system size becoming
too large or too small.

Studies using the Lennard--Jones potential showed the calculation
progressing through the FCC local minimum and two other stacking--fault
local minima before finding the HCP global minimum enthalpy structure.
This was shown to be repeatable and efficient.

When this new GA was used to study phases of the Dzugutov potential
at different pressures all the previously reported zero--temperature
phases were found, along with three new phases, one of which is the
most stable phase at all positive pressures. These new structures
are markedly different from the three previously--known phases. This
clearly illustrates the power of this GA to find new crystal structures
that were hitherto unexpected.

\section{Acknowledgments}

Calculations were performed on our departmental Beowulf cluster, EPSRC
grant R47769 from the Multi-Project Research Equipment Initiative.
NLA is grateful to the EPSRC for financial support. The authors acknowledge useful discussions with Peter Main.

\bibliographystyle{apsrev}

\end{document}